\documentclass[paper]{JHEP3}
\usepackage{yfonts}[1988/10/03]
\newcommand {\beq}{\begin{equation}}
\newcommand {\eeq}{\end{equation}}
\newcommand {\beqa}{\begin{eqnarray}}
\newcommand {\eeqa}{\end{eqnarray}}
\newcommand {\n}{\nonumber \\}
\newcommand {\del}{\partial}
\newcommand{\g}{\textfrak{g}}

\title{Density-metric unimodular gravity: vacuum maximal symmetry} 
\author{Amir H. Abbassi\\
Department of Physics, School of Sciences, Tarbiat Modares
University,\\ P.O.Box 14155-4838, Tehran, Iran.\\
E-mail:\email{ahabbasi@modares.ac.ir}}
\author{Amir M. Abbassi\\
Department of Physics, University of Tehran,\\ P.O.Box 14155-6455, Tehran, Iran.\\
E-mail:\email{amabasi@khayam.ut.ac.ir}}
\abstract{We have investigated the vacuum maximally symmetric solutions of recently proposed density-metric unimodular gravity theory, the results are widely different from inflationary scenario. The exponential dependence on time in deSitter space is substituted by a power law.  Open space-times with non-zero cosmological constant are excluded .}
\preprint{***}
\begin{document}

%%%%%%%%%%%%%%%%%%%%%%%%%%%%%%

\section{Introduction}
Unimodular gravity is a very modest modification of Einstein general theory of relativity
which leads to the same field equations with the cosmological constant emerging as an integration
constant. In unimodular theory one apparently reduces the dynamical components of the metric by one. This
is accomplished by discharging the determinant of the metric as dynamical variable.
Conventionally this demand is granted by imposing $|g|=1$ as a constraint[1-9].
However, it was shown that one indeed recovers the same number of dynamical 
degrees of freedom as in GR via a tertiary constraint[10]. Even though a covariant expression on variation of the metric, $g_{\mu\nu}\delta g_{\mu\nu}=0$ emerges from the unitary condition $|g|=1$, but the primary relation itself is not established on a covariant form.
Determinant of the metric is a scalar density and cannot be kept equal to one in each arbitrary coordinate system. Evidently, this is not compatible with our notion of general covariance. This disadvantage has been removed in density-metric unimodular gravity proposed recently[11]. In metric theories of gravity the gravitational field is represented by the spacetime metric which is a symmetric tensor field of rank $(0,2)$ and signature of $+2$ defined on a four dimensional manifold. In density-metric theory the gravitational field is represented by a symmetric tensor density field of rank $(0,2)$ and weight $-\frac 12$ which its determinant is equal to one by definition. Since the determinant of a tensor density of weight $-\frac12$ in a four dimensional spacetime is a scalar under coordinate transformations, so the unitary condition may be imposed consistently with the principle of general covariance. Using this density metric we may equip the manifold by a specific connection.  
It turns out that introducing torsion-free
and metric compatibility conditions are not sufficient to fix uniquely all components of the connection.
Actually, the components of trace of connection remain undetermined. Hence the suppositions are:
\beqa
& & \g_{\mu\nu}=\g_{\nu\mu}\;\;\;\;\;{\mbox{\small Symmetric tensor density of weight}-\frac12} \label{a1}\\
& & |\g|=1\;\;\;\;\;\;\;\;\; {\mbox{\small Unitary condition}}\label{a2}\\
& & \Gamma^\rho_{\mu\nu}=\Gamma^\rho_{\nu\mu}\;\;\; {\mbox{\small Torsion free}}\label{a3}\\
& & \nabla_\rho\g_{\mu\nu}=0\;\;\; {\mbox{\small Metric compatible}}\label{a4}
\eeqa 
By expanding out the equation (\ref{a4}) for three different permutations of indices
and subtracting the second and third of these form the first we obtain:
\beqa\label{a5}
\Gamma^\lambda_{\mu\nu}=&\frac12&\g^{\lambda\rho}(\del_\mu\g_{\nu\rho}
+\del_\nu\g_{\mu\rho}-\del_\rho\g_{\mu\nu})\n 
&+&\frac14(\Gamma^\rho_{\rho\mu}\delta^\lambda_\nu+\Gamma^\rho_{\rho\nu}\delta^\lambda_\mu
-\Gamma^\rho_{\rho\kappa}\g^{\kappa\lambda}\g_{\mu\nu})
\eeqa
Contraction of indices $\lambda$ and $\mu$ in (\ref{a5}) gives
\beq \label{a5-1}
\Gamma^\lambda_{\lambda\nu}=\frac12 \g^{\lambda\rho}\partial_\nu\g_{\lambda\rho}
+\Gamma^\rho_{\rho\nu}
\eeq
Obviously (\ref{a5-1}) brings us to
\beq \label{a5-2}
\g^{\lambda\rho}\partial_\nu\g_{\lambda\rho}=0
\eeq
So four relations of (\ref{a5-2}) reduce ten independent components of density-metric to six. 
Therefore components of connection typically are functions of six independent components of density-metric and four independent components of the trace of connection $\Gamma^\rho_{\rho\mu}$.
%The mass distribution should have three equal principal moments of inertia. Here %$C^\prime$ is a
%good candidate to be proportional to this principal moment of inertia of the source %term.
Using the connection(\ref{a5}), we may introduce Riemann tensors and Ricci tensor
and scalar respectively by:
\beqa
R^\rho_{\;\sigma\mu\nu}&=&\del_\nu\Gamma^\rho_{\mu\sigma}-\del_\mu\Gamma^\rho_{\nu\sigma}
+\Gamma^\rho_{\nu\lambda}\Gamma^\lambda_{\mu\sigma}-\Gamma^\rho_{\mu\lambda}
\Gamma^\lambda_{\nu\sigma}\label{a6}\\
R_{\rho\sigma\mu\nu}&=&\g_{\rho\lambda}R^\lambda_{\;\sigma\mu\nu}\label{a7}\\
R_{\sigma\nu}&=&R^\rho_{\;\sigma\rho\nu}\label{a8}\\
R&=&\g^{\sigma\nu}R_{\sigma\nu}\label{a9}
\eeqa
Here (\ref{a6}) is a tensor of rank $(1,3)$, (\ref{a7}) is a tensor density of rank
$(0,4)$ and weight $-\frac12$, (\ref{a8}) is a tensor of rank $(0,2)$ and (\ref{a9})
is a scalar density of weight $+\frac12$. Let us notice some algebraic properties of Riemann tensor. It is antisymmetric in its first two
indices 
\beq \label{a10}
R_{\rho\sigma\mu\nu}=-R_{\rho\sigma\nu\mu}
\eeq
and it is not invariant under interchange of the first pair of indices with the second:
\beq\label{a11}
R_{\rho\sigma\mu\nu}\neq R_{\mu\nu\rho\sigma}
\eeq
The sum of the cyclic permutations of the last three indices of Riemann tensor vanishes:
\beq\label{a12}
R_{\rho\sigma\mu\nu}+R_{\rho\mu\nu\sigma}+R_{\rho\nu\sigma\mu}=0
\eeq
The Bianchi identity holds too. The sum of cyclic permutation of the first three indices of covariant derivatives of the Riemann tensor is zero:
\beq\label{a13}
R_{\rho\sigma\nu;\lambda}+R_{\rho\sigma\lambda\mu;\nu}
+R_{\rho\sigma\nu\lambda;\mu}=0
\eeq
The Ricci tensor associated with the Christoffel connection is symmetric but in our case this is not generally true. Twice contraction of Bianchi identity (\ref{a13}) with $\g^{\rho\mu}$
and $\g^{\sigma\nu}$ plus doing some calculations in Riemann normal
coordinates result in:
\beq\label{a14}
\nabla^\sigma(\frac14R_{\lambda\sigma}+\frac34R_{\sigma\lambda}-\frac12
\g_{\sigma\lambda}R)=0
\eeq
In density-metric unimodular gravity figured that $R^2$ was the simplest possible choice for 
a Lagrangian density and proposed
\beq\label{a15}
I=\int{\kappa R^2dx^4}
\eeq
for the action of gravitational fields in the absence of matter and other kinds of sources.
Variation of the action (\ref{a15}) with respect to $\Gamma^\rho_{\rho\lambda}$
leads to the field equation
\beq\label{a16}
\nabla_\lambda R=0
\eeq
Also variations of the action (\ref{a15}) with respect to $\g^{\mu\nu}$ and applying the 
unimodular condition
\beq\label{a17}
\g_{\mu\nu}\delta\g^{\mu\nu}=0
\eeq
by the method of Lagrange undetermined multipliers and inserting field equation
(\ref{a16}) lead to
\beq\label{a18}
R_{[\mu\nu]}-\frac14 \g_{\mu\nu}R=0
\eeq
 We should mention that density-metric is not eligible to carry out the whole tasks of a common metric. First its operation on two vectors gives a scalar density instead of a scalar. Second the density-metric has six independent components. In accordance with affine theories of gravity we may equip the spacetime with a metric by defining
\beq
g_{\mu\nu}=R\g_{\mu\nu} \label{a19}
\eeq
where $R$ is given by eq.(\ref{a9}) and of course should be nonzero. In addition it depends on four independent components of trace of connection,$\Gamma^\rho_{\rho\mu}$. Thus $g_{\mu\nu}$ depends on ten independent dynamical variables. Given the equation (\ref{a19}) in principle, we can work out density-metric and trace of connection in terms of ten independent components of $g_{\mu\nu}$ and formally write:
\beq
\g_{\mu\nu}=\g_{\mu\nu}(g_{\alpha\beta}) \;\;,\;\;\Gamma^\rho_{\rho\mu}=\Gamma^\rho_{\rho\mu}(g_{\alpha\beta}) \label{a20}
\eeq 
We have already worked out the solution of this theory for vacuum spherically symmetric space-time:
\beq\label{a21}
ds^2=B(r)dt^2-A(r)dr^2-r^2(d\theta^2+sin^2\theta d\phi^2)
\eeq
with
\beq\label{a22}
B(r)=(1+\frac{C^\prime}{r^3})^{-\frac23}[1-\frac{2GM-\Lambda C^\prime}{r}
(1+\frac{C^\prime}{r^3})^{-\frac13}
+\Lambda r^2(1+\frac{C^\prime}{r^3})^{-\frac13}]
\eeq 
and
\beq\label{a23}
A(r)=(1+\frac{C^\prime}{r^3})^{-2}[1-\frac{2GM-\Lambda C^\prime}{r}(1+\frac{C^\prime}{r^3})
^{-1}+\Lambda r^2(1+\frac{C^\prime}{r^3})^{-\frac13}]^{-1}
\eeq
In these relations $M$ is the mass of point object which is the source of spherically symmetry and $\Lambda$ has dimensions$[L^{-2}]$ where may be identified as cosmological constant.
$C^\prime$ is a new constant which appears in this theory with dimensions$[L^{-3}]$.\\
The constant $C^\prime$ has a physical 
interpretation, a convenient way to show its real nature is when we manipulate 
the post-post-Newtonian formalism ; just as the same way that the Schwarzschild
radius is fixed by comparison with Newtonian limits. It should be noticed that a 
spherically symmetric matter distribution has a total mass, but certainly
has a vanishing quadrupole moment. Such a matter distribution could have three equal principal moments of inertia $I$. It seems very probable that  $C^\prime\propto\frac{GI}{c^2}$. In this way $C^\prime$ is an individual property of the source object.  
Observational data from double pulsar systems can be used to verify this issue.
Here we can offer an upper bound for it.\\       
In the next section we explicitly work out these results and correctly predict what are observed in three classical tests of gravitation.
    
\section{Classical Tests}
For this purpose we may choose $\Lambda=0$. The geodesic equations imply that the quantity
\beq\label{b1}
\epsilon=-g_{\mu\nu}\frac{dx^\mu}{d\lambda}\frac{dx^\nu}{d\lambda}
\eeq
is constant along the path. For massive particles we typically choose $\lambda$ so that $\epsilon=1$ . While for massless particles we always have $\epsilon=0$.
Invariance under time translations leads to conservation of energy while invariance under spatial rotations leads to conservation of the three components of angular momentum. Conservation of direction of angular momentum means that particle moves on a plane. So we may choose $\theta=\frac{\pi}{2}$. The two remaining Killing vectors correspond to the energy and magnitude of angular momentum. The energy arises from time-like Killing vector $(\partial_t)^\mu=(1,0,0,0)$ while the Killing vector whose conserved quantity is the magnitude of angular momentum is $(\partial_\phi)^\mu=(0,0,0,1)$. The two conserved quantities are
\beq\label{b2}
E=-(1+\frac{C^\prime}{r^3})^{-\frac 23}[1-\frac{2GM}{r}(1+\frac{C^\prime}{r^3})^{-\frac 13}]\frac{dt}{d\lambda}
\eeq
and
\beq\label{b3}
L=r^2\frac{d\phi}{d\lambda}
\eeq
From eq.(\ref{b1})by using eqs.(\ref{b2}) and (\ref{b3}) we can arrive at the following
result:
\beqa\label{b4}
&&-E^2+(1+\frac{C^\prime}{r^3})^{-\frac 83}(\frac{dr}{d\lambda})^2+\n
&&(1+\frac{C^\prime}{r^3})^{-\frac 23}(1-\frac{2GM}{r}(1+\frac{C^\prime}{r^3})^{-\frac 13})(\frac{L^2}{r^2}+\epsilon)=0\quad
\eeqa
Then expanding in powers of $r$, we shall content ourselves here with only the lowest order of approximation in eq.(\ref{b4}). So we have
\beq\label{b5}
(\frac{dr}{d\lambda})^2+(1-\frac{2GM}{r})(\frac{L^2}{r^2}+\epsilon)-
E^2(1+\frac 83\frac{C^\prime}{r^3})+\frac{2\epsilon C^\prime}{r^3}=0
\eeq
Now we are prepared to discuss the classical tests one by one.\\
1){\it Precession of Perihelia}: With $\epsilon=1$ in eq.(\ref{b5}) and multiplying
it by $(\frac{d\phi}{d\lambda})^{-2}=\frac{r^4}{L^2}$ and defining
$x=\frac{L^2}{GMr}$ we have
\beq
(\frac{dx}{d\phi})^2-2x+x^2-\frac{2G^2M^2}{L^2}x^3+\frac{GM}{L^4}
(2C^\prime-\frac{8E^2C^\prime}{3})x^3\n
=\frac{E^2L^2}{G^2M^2}-\frac{L^2}{G^2M^2}\label{b6}
\eeq
Differentiating eq.(\ref{b6}) with respect to $\phi$ gives
\beq\label{b7}
\frac{d^2x}{d\phi^2}-1+x=(\frac{3G^2M}{L^2}-\frac{3GM}{L^4}C^\prime
+\frac{4GM E^2C^\prime}{L^4})x^2
\eeq
After some manipulation [12] we obtain that the perihelion advances by an angle
\beq\label{b8}
\Delta\phi=\frac{6\pi G^2M^2}{L^2}(1-\frac{C^\prime}{GML^2}+\frac{4C^\prime E^2}{3GML^2})
\eeq
On the other hand we have
\beq\label{b9}
E^2=1+\frac{G^2M^2}{L^2}(e^2-1)\quad{\rm and}\quad L^2\cong GM(1-e^2)a
\eeq
where $e$ is the eccentricity and $a$ is the semi-major axis of the ellipse.
Putting eq.(\ref{b9}) in eq.(\ref{b8}) yields
\beq\label{b10}
\Delta\phi=\frac{6\pi GM}{(1-e^2)a}(1+\frac{C^\prime}{3G^2M^2(1-e^2)a}
+\frac{4C^\prime}{3GM(1-e^2)a^2})
\eeq
The third term in the parenthesis is smaller than the second term by a 
$\frac{GM}{a}$ factor and may be neglected.
\beq\label{b11}
\Delta\phi=\frac{6\pi GM}{(1-e^2)a}(1+\frac{C^\prime}{3G^2M^2(1-e^2)a})
\eeq
Observational data put an upper bound on the $C^\prime$.\\
 2){\it Deflection of Light}: Putting $\epsilon =0$ in eq.(\ref{b5}) and multiplying it by $(\frac{d\phi}{d\lambda})^{-2}=\frac{r^4}{L^2}$ and defining
$u=\frac 1r$ we have
\beq\label{b12}
(\frac{du}{d\phi})^2+u^2-2GMu^3=\frac{E^2}{L^2}(1+8C^\prime u^3)
\eeq
Differentiation of eq.(\ref{b12}) with respect to $\phi$ makes
\beq\label{b13}
\frac{d^2u}{d\phi^2}+u=(3GM+\frac{12C^\prime E^2}{L^2})u^2
\eeq
By some straight forward calculations [13]  the deflection angle is
\beq\label{b14}
\delta=\frac 4{r_0}[GM+\frac{4C^\prime E^2}{L^2}]
\eeq
where $r_0$ is the distance of closest approach to the origin.
Since $\frac{E}{L}$ is $\frac 1{r_0}$ then eq.(\ref{b14}) becomes
\beq\label{b15}
\delta=\frac{4GM}{r_0}+\frac{16C^\prime}{{r_0}^3}
\eeq
An upper bound on $C^\prime$ may be obtained by comparison with observational data.\\ 
3){\it Gravitational Redshift}: Let us Consider an observer with four-velocity $U^\mu$ who is stationary $(U^i=0)$, we have:
\beq\label{b16}
U^0=(B(r))^{-\frac 12}
\eeq
Any such observer measures the frequency of a photon following along a null geodesic $\chi^\mu (\lambda)$ to be
\beq\label{b17}
\omega=-g_{\mu\nu}\frac{dx^\nu}{d\lambda}U^\mu={B(r)}^{\frac 12}\frac{dt}{d\lambda}
\eeq
By considering eq.(\ref{b2}) and (\ref{b2}) we conclude that for a photon emitted at $r_1$
and observed at $r_2$ the observed and emitted frequencies are
related by
\beq\label{b18}
\frac{\omega_2}{\omega_1}=(\frac{B(r_1)}{B(r_2)})^{\frac 12}
\eeq
In the interval $\Lambda^{-\frac 12}\gg r\gg 2GM$ and ${C^\prime}^{-\frac 13}$
we get
\beq\label{b19}
\frac{\omega_2}{\omega_1}=1-GM(\frac 1{r_1}-\frac 1{r_2})-\frac{C^\prime}{3}
(\frac 1{{r_1}^3}-\frac 1{{r_2}^3})+\frac \Lambda 2(r_1^2-r_2^2)
\eeq
This is also consistent with observational constraints.\\

Now we want to work out the solutions of these field equations for the vacuum maximal symmetry space-times. In other words our goal is to illustrate the analogue of Minkowski , deSitter and anti- deSitter space-times in the density metric unimodular theory of gravity. The deSitter-Schwarzschild analogue in static form may be obtained by putting $M=C^\prime=0$ in eqs.(\ref{a22})and
(\ref{a23}).   

\section{Maximal Symmetry}
A rigorous mathematical analysis may be applied by taking $N$ dimensional
manifold equipped with a density metric and a trace of connection. Assuming the 
density metric of rank $(0,2)$ and weight $-\frac12$ being form invariant
under a given coordinate transformation generated by a Killing vector $\xi^\mu$ 
leads to:
\beq\label{c1}
\nabla_\mu\xi_\nu+\nabla_\nu\xi_\mu-\frac12 \g_{\mu\nu}\nabla_\lambda\xi^\lambda=0
\eeq
This means that $\nabla_\mu\xi_\nu-\frac14 \g_{\mu\nu}\nabla_\lambda\xi^\lambda$
is an anti-symmetric tensor. Since the trace of connection is taken as dynamical variable the condition to be form invariant provide us:
\beq\label{c2}
\nabla_\nu\nabla_\lambda\xi^\lambda=\xi^\lambda(\frac{\del}{\del x^\nu}\Gamma^\rho
_{\rho\lambda}-\frac{\del}{\del x^\lambda}\Gamma^\rho_{\rho\nu})\equiv
{\hat R}_{\nu\lambda}\xi^\lambda
\eeq
It can be easily shown that
\beq\label{c3}
\nabla_\sigma\nabla_\rho\xi_\mu=-R^\lambda_{\sigma\rho\mu}\xi_\lambda
-\frac14 \g_{\sigma\rho}{\hat R}_{\mu\lambda}\xi^\lambda
+\frac14 \g_{\sigma\mu}{\hat R}_{\rho\lambda}\xi^\lambda
+\frac14 \g_{\mu\rho}{\hat R}_{\sigma\lambda}\xi^\lambda
\eeq
Following the same argument as given by Weinberg[14], we may conclude that the 
manifold should have $\frac{N(N+1)}{2}$ Killing vectors to be maximally symmetric. We start from general formula for commutator of covariant derivatives of tensors
\beqa
& &\nabla_\sigma\nabla_\nu(\nabla_\mu\xi_\rho-\frac14 \g_{\mu\rho}\nabla_\eta\xi^\eta)
-\nabla_\nu\nabla_\sigma(\nabla_\mu\xi_\rho-\frac14 \g_{\mu\rho}\nabla_\eta\xi^\eta)=\n
& &-R^\lambda_{\rho\sigma\nu}(\nabla_\mu\xi_\lambda-\frac14 \g_{\lambda\mu}\nabla_\eta\xi^\eta)-R^\lambda_{\mu\sigma\nu}(\nabla_\lambda\xi_\rho
-\frac14 \g_{\lambda\rho}\nabla_\eta\xi^\eta)\label{c4}
\eeqa 
Then apply eqs.(\ref{c1}),(\ref{c2}) and (\ref{c3}) into eq.(\ref{c4}). 
It can be shown that for a maximally symmetric manifold we should have:
\beq\label{c5}
{\hat R}_{\mu\nu}=0
\eeq
This is an important result, since it causes the Riemann curvature tensor to resume
the same algebraic property just as Riemann geometry, i.e. if Christoffel connections were used. Following Weinberg's procedure we obtain
\beq\label{c6}
(N-1)R_{\lambda\rho\sigma\nu}=R_{\nu\rho}\g_{\lambda\sigma}-R_{\sigma\rho}
\g_{\lambda\nu}
\eeq
and
\beq\label{c7}
R_{\sigma\rho}=\frac 1N \g_{\sigma\rho}R
\eeq
Putting (\ref{c7}) into (\ref{a14}) gives
\beq\label{c8}
\nabla_\lambda R=0
\eeq
Of course this does not mean $R$ is constant since it is a density scalar.  

\section{Vacuum Cosmic Dynamics}
The homogeneity and isotropy of a space-time imply that it is maximally symmetric. In co-moving coordinates the metrics are given by Robertson-Walker line element,
\beq\label{d1}
ds^2\ =dt^2-a^2(t)\left\{\frac{dr^2}{1-k r^2} +r^2(d\theta^2+\sin^2\theta d\phi^2)\right\}
\eeq  
To prepare the appropriate density metric consider the following scenario: Multiply the components of Robertson-Walker metric by inverse of$\sqrt[4]{|g|}$ , where$|g|$ is determinant of the metric. This results in a density tensor of rank (0,2) with the desired weight $-\frac 12$ . The nonzero components of the new density metric are,
 \beqa \label{d2}
\g_{tt}&=&-\frac{(1-k r^2)^{\frac 14}}{a^{\frac 32} r\sin^{\frac 12}\theta}\;\;\;, \; \g_{rr}=\frac{a^{\frac 12}}{r\sin^{\frac 12}\theta (1-k r^2)^{\frac 34}}\;\;\;\;\;\;\; \n 
\g_{\theta\theta}&=&\frac{a^{\frac 12} r(1-k r^2)^{\frac 14}}{\sin^{\frac 12}\theta} , \g_{\phi\phi}= a^{\frac 12} r\sin^{\frac 12}\theta (1-k r^2)^{\frac 14}
\eeqa                                    
and the nonzero components of its inverse are,
 \beqa  \label{d3}
 \g^{tt}=-\frac{a^{\frac 32} r\sin^{\frac 12}\theta}{(1-k r^2)^{\frac 14}} \; &,&\; \g^{rr}=\frac{r\sin^{\frac 12}\theta (1-k r^2)^{\frac 34}}{a^{\frac 12}}\;\;\;\;\;\;\;\;\;\;\;\; \n
\g^{\theta\theta}=\frac{\sin^{\frac 12}\theta}{a^{\frac 12}r(1-k r^2)^{\frac 14}} &,& \g^{\phi\phi}=\frac 1{a^{\frac 12} r\sin^{\frac 32}\theta (1-k r^2)^{\frac 14}}
 \eeqa 
In terms of the unknown components of trace of connection, the nonzero components of the connection are determined by using eqs.(\ref{d2}),(\ref{d3}) and (\ref{a5}).
The results are presented in A.1.
 The off diagonal components of the field equation (\ref{a8}) simply are:
 \beq \label{d4}
 R_{\mu\nu}=\;0 \qquad for\;\;\;\mu\neq\nu
 \eeq                                                                   
Inserting the connection from A.1 into eqs.(\ref{a6}), and (\ref{a8}), then
$R_{rt}=0$ gives
 \beq\label{d5}
 (\Gamma^\rho_{\rho t}+\frac{\dot a}{a})(\Gamma^\rho_{\rho r}-\frac 2r
 -\frac{kr}{2(1-kr^2)})=0
 \eeq                                                                
Equation (\ref{d5}) derives a manifestly unique expression for the radial component of trace of connections.
 \beq \label{d6}
 \Gamma^\rho_{\rho r}=\frac 2r+\frac{kr}{1-kr^2}
 \eeq                                                                  
  Similarly the polar component of trace of connection will be obtained by considering the equation $R_{t\theta}=0$. It gives :
 \beq\label{d7}
(\Gamma^\rho_{\rho t}+\frac{\dot a}{a})(\Gamma^\rho_{\rho\theta}-\cot\theta)
-2 \frac{\partial}{\partial t}\Gamma^\rho_{\rho\theta}=0
 \eeq                                                     
Eq. (\ref{d7}) imposes a unique expression for polar component of connection trace
\beq \label{d8}
\Gamma^\rho_{\rho\theta}=\cot\theta
\eeq                     
 The azimuth component of trace of connection will emerge immediately by considering $R_{t\phi}=0$ . This simply reads
\beq \label{d9}
\Gamma^\rho_{\rho\phi}(\Gamma^\rho_{\rho t}+\frac{\dot a}{a})
=0
\eeq                             
Therefore eq.(\ref{d9}) at last determines azimuth component of connection trace
\beq \label{d10}
\Gamma^\rho_{\rho\phi}=0
\eeq           
The other off diagonal components of field equations, $R_{r\theta}=R_{r\phi}=R_{\theta\phi}=0 $ ,
are satisfied automatically by the given equations(\ref{d6}),(\ref{d8})and(\ref{d10}).
Plugging equations (\ref{d6}),(\ref{d8})and (\ref{d10}) into the given components of the connection in A.1 reduce the non-vanishing components of the connection.
They are given in A.2. These results may be used to compute the diagonal components
of Ricci tensor:
 \beqa
 R_{tt}&=&\frac 34(\frac{\ddot a}{a}+\frac{\dot a}{a}\Gamma^\rho_{\rho t}+
 \dot{\Gamma}^\rho_{\rho t}) \label{21} \label{d11}\\
 R_{rr}&=&-\frac{2k}{1-kr^2}-\frac{a^2}{4(1-kr^2)}[\frac{\ddot a}{a}+\frac 12
 (\frac{\dot a}{a})^2+2\frac{\dot a}{a}\Gamma^\rho_{\rho t}+\dot{\Gamma}^\rho_{\rho t}+\frac 12{\Gamma^\rho_{\rho t}}^2] \label{d12}\\
 R_{\theta\theta}&=&-2kr^2-\frac 18a^2r^2[2\frac{\ddot a}{a}+(\frac{\dot a}{a})^2+2\dot{\Gamma}^\rho_{\rho t}+4\frac{\dot a}{a}\Gamma^\rho_{\rho t} 
 +{\Gamma^\rho_{\rho t}}^2] \label{23} \label{d13}\\
 R_{\phi\phi}&=&R_{\theta\theta}\sin^2\theta \label{d14}                    
 \eeqa                                                             
Let us write the diagonal components of the field equation 
(\ref{a18}) :
\beqa 
\frac 32\g^{tt}R_{tt}-\frac 12\g^{rr}R_{rr}-\g^{\theta\theta}R_{\theta\theta}
&=&0 \label{25} \label{d15}\\
\frac 32\g^{rr}R_{rr}-\frac 12\g^{tt}R_{tt}-\g^{\theta\theta}R_{\theta\theta}
&=&0\label{26} \label{d16}\\
\g^{\theta\theta}R_{\theta\theta}-\frac 12\g^{tt}R_{tt}-\frac 12\g^{rr}R_{rr}
&=&0 \label{d17} 
\eeqa                                                    
The azimuth component of eq.(\ref{a18}) reproduces the same equation as eq.(\ref{d17}).
To achieve the final result it is sufficient to substitute eqs. (\ref{d11})-(\ref{d13}) into eqs. (\ref{d15})-(\ref{d17})  which all merely get to the same equation as,
 \beq \label{d18}
 \frac{\ddot a}{a}-\frac 14(\frac{\dot a}{a})^2-\frac{4k}{a^2}+\frac 12
 \frac{\dot a}{a}\Gamma^\rho_{\rho t}+\dot{\Gamma}^\rho_{\rho t}
 -\frac 14{\Gamma^\rho_{\rho t}}^2=0
 \eeq                                                                   
The field equation (\ref{a16}) does not express a new equation and indeed relation (\ref{d18}) already satisfies equation (\ref{a16}). In the case of Minkowski space-time, $a$ ,is constant and $k$ is equal to zero. These are in agreement with relation (\ref{d18}) if we distinguish,
 \beq\label{d19}
 \Gamma^\rho_{\rho t}=0
 \eeq                                                         
According to arguments expounded in [11],concerning gauge fixing,
we regard that generally relation (\ref{d19}) rigorously holds. This is consistent with our result for vacuum spherical symmetry and $M=0$ presented in the mentioned reference. Finally substituting relation (\ref{d19}) into eq.(\ref{d18}) yields 
 \beq\label{d20}
 \frac{\ddot a}{a}-\frac 14(\frac{\dot a}{a})^2-\frac{4k}{a^2}=0
 \eeq                                                                            
It is relevant to compare the obtained equation (\ref{d20}) with its counterpart in general theory of relativity which is Friedmann equation. We have [15],
 \beq\label{d21}
 {\dot a}^2=\frac{\Lambda}{3}a^2-k
 \eeq
 \beq\label{d22}
 2\frac{\ddot a}{a}+(\frac{\dot a}{a})^2+\frac{k}{a^2}-\Lambda=0
 \eeq                                                                           
Eliminating $\Lambda$ between equations (\ref{d21}) and(\ref{d22}) we get,
\beq \label{d23}
\frac{\ddot a}{a}-(\frac{\dot a}{a})^2-\frac{k}{a^2}=0
\eeq  
It is interesting to notice that in eqs. (\ref{d20}) and (\ref{d23}) identical terms appear with merely different numerical coefficients and both equations are invariant under time reversal transformation. So it is plausible while they have probably the same roots we predict to obtain results with different behavior than the existing models. We are going to solve the eq.(\ref{d20}). To do this task,we assume in accordance with eq.(\ref{d21})
it has a solution of general form
\beq\label{d24}
{\dot a}^2=f(a)
\eeq
Differentiating eq.(\ref{d24}) with respect to time and taking ${\dot a}\neq0$
yields
\beq\label{d25}
\ddot{a}=\frac 12\frac{df(a)}{da}
\eeq
Plugging eqs.(\ref{d24}) and (\ref{d25}) into eq.(\ref{d20}) gives 
\beq\label{d26}
\frac a2\frac{df(a)}{da}-\frac 14 f(a)-4k=0
\eeq
Eq.(\ref{d24}) has the general solution,
\beq\label{d27}
{\dot a}^2\equiv f(a)=\Lambda\sqrt{a}-16k
\eeq
where $\Lambda$ is constant of integration and we have intended to call it so 
to show its resemblance with cosmological constant in Friedmann model.
Eq.(\ref{d27}) may be integrated and has the general solution 
\beq\label{d28}
\pm t=\frac{4(\Lambda\sqrt{a}+32k)}{3\Lambda^2}\sqrt{\Lambda\sqrt{a}-16k}+C
\eeq
where C is the second constant of integration and may be put equal to
zero without loose of generality. To find scale factor as a function of time from
eq.(\ref{d28}) we will face with a cubic equation. In the next section we will
discuss some features of the solutions of this equation in different situations.

\section{Empty Cosmological Models}
Reality condition imposed on (\ref{d27}) and (\ref{d28}) imply the following limits on the domains of $a$ and $t$ without solving 
the equations.
\beqa
I\hspace{1 cm}\Lambda>0\hspace{1.0 cm}k&=&1\hspace{2 cm}a\geq(\frac{16}{\Lambda})^2\hspace{1cm}t\in R\hspace{1.3 cm}\mbox{deSitter}
 \label{e1}\\
II\hspace{1 cm}\Lambda>0\hspace{1.0 cm}k&=&0\hspace{2 cm} a\geq0\hspace{1.7 cm}t\in R\hspace{1.3 cm}\mbox{deSitter}
 \label{e2}\\ 
III\hspace{1cm}\Lambda>0\hspace{1.0 cm} k&=&-1\hspace{1.7cm}a\geq0\hspace{1,5cm}|t|\geq\frac23(\frac{16}{\Lambda})^2\hspace{.4cm}\mbox{deSitter}\label{e3} \\ 
IV\hspace{1cm} \Lambda=0\hspace{1.0cm}k&=&1\hspace{3cm}\mbox{no solution}\label{e4}\\
V\hspace{1cm}\Lambda=0\hspace{1.0cm}k&=&0\hspace{2cm}a=const\hspace{1cm}
t\in R\hspace{.9cm}\mbox{Minkowski}\label{e5}\\
VI\hspace{1cm}\Lambda=0\hspace{1.0cm}k&=&-1\hspace{1.7cm}a\propto t\hspace{1.8cm}
t\in R\hspace{1.7cm}\mbox{Milne}\label{e6}\\ 
VII\hspace{1cm}\Lambda<0\hspace{1.0cm}k&=&+1\hspace{2.8cm}\mbox{no solution}\label{e7}\\
VIII\hspace{1cm}\Lambda<0\hspace{1.0cm}k&=&0\hspace{3cm}\mbox{no solution}\label{e8}\\
IX\hspace{1cm}\Lambda<0\hspace{1.0cm}k&=&-1\hspace{0.8cm}0\leq a\leq(\frac{16}{\Lambda})^2
\hspace{.3cm}|t|\leq\frac 23(\frac{16}{\Lambda})^2\hspace{0.2cm}\mbox{anti-deSitter}\label{e9}
\eeqa
Now we express some points for each case in more detail.
\begin{itemize}
\item[I.]{\bf $\Lambda>0,k=1$}\\
All closed models with positive lambda have a lower bound on $a$
equal to $a_{min}=\frac{256}{\Lambda^2}$. To find $a$ as a function of time from
eq.(\ref{d28}) we face with a cubic equation which its discriminant is positive for 
this model. This means it merely has one real root. it is:
\beqa\label{e10}
a&=&\{\frac{16}{\Lambda}[\sqrt[3]{1+9(\frac{\Lambda}{16})^4t^2+
\sqrt{-1+(1+9(\frac{\Lambda}{16})^4t^2)^2}}\n
&+&\sqrt[3]{1+9(\frac{\Lambda}{16})^4t^2-\sqrt{-1+(1+9(\frac{\Lambda}{16})^4t^2)^2}}
-1]\}^2
\eeqa
Models of class I possess event horizon which can be find by considering
\beq\label{e11}
\int^{r_H}_{0}{\frac{dr}{\sqrt{1-r^2}}}=\int^{+\infty}_{t_0}{\frac{dt}{a(t)}}
\eeq
We infer from eq.(\ref{e10}) that $a\rightarrow +\infty$ as $t\rightarrow +\infty$.
Taking $a(t_0)=a_0$, eq.(\ref{e11}) gives
\beq\label{e12}
sin^{-1}(r_H)=\int^{+\infty}_{t_0}{\frac{da}{a\dot{a}}}
\eeq
Calculating the integral in eq.(\ref{e12}) by inserting eq.(\ref{d27}),
we will have
\beq\label{e13}
r_H=cos(tan^{-1}\sqrt{\frac{\Lambda\sqrt{a_0}}{16}-1})
\eeq
\item[II.] {\bf $\Lambda>0,k=0$}\\
In flat models with positive lambda eq.(\ref{d27}) reduces to
\beq\label{e14}
\frac{\dot{a}}{\sqrt[4]{a}}=\pm\sqrt{\Lambda}
\eeq
and integration of eq.(\ref{e14}) with respect to time gives
\beq\label{e15}
a=(\frac 34\sqrt{\Lambda}|t|)^{\frac43}
\eeq
The integration constant has been chosen so that we have
$a=0$ at $t=0$. While Hubble and deceleration parameters in Friedmann empty
flat model are constant, eq.(\ref{e15}) implies
\beq\label{e16}
H\equiv \frac{\dot a}{a}=\frac 4{3t}
\eeq
and
\beq\label{e17}
q\equiv-\frac{\ddot a}{aH^2}=-\frac 1{4t^2}
\eeq
Deceleration parameter eq.(\ref{e17}) stands for an accelerating Universe
which its acceleration rate decreases by passing of time.\\
The exponential growth of deSitter space in models $I$ and $II$ is replaced by a power law 
which practically is more plausible to match the physical reality.
This behavior is in no way inflationary.
Model $II$ also possesses an event horizon
\beq\label{e18}
r_H=\int^{+\infty}_{t_0}{\frac{dt}{a(t)}}=\int^{+\infty}_{t_0}
{\frac 1{a_0}(\frac{t_0}{t})^\frac 43 dt}=\frac 1{3a_0}
\eeq
where $a_0=a(t_0)$. It should be noted this model is singular at $t=0$.
\item[III.] {\bf $\Lambda>0, k=-1$}\\
Eq.(\ref{d28}) for an open model with positive lambda has a real root
\beqa\label{e19}
a&=&\{\frac{16}{\Lambda}[\sqrt[3]{-1+9(\frac{\Lambda}{16})^4t^2+
\sqrt{-1+(-1+9(\frac{\Lambda}{16})^4t^2)^2}}\n
&+&\sqrt[3]{-1+9(\frac{\Lambda}{16})^4t^2-\sqrt{-1+(-1+9(\frac{\Lambda}{16})^4t^2)^2}}
-1]\}^2
\eeqa
which is defined merely for $|t|>\frac 23(\frac{16}{\Lambda})^2$. While it is singular at 
$t=\frac 23 (\frac{16}{\Lambda})^2$ is not defined for long intervals of time.
For this reason this model can not present a physically acceptable space-time.
\item[IV.]{\bf$\Lambda=0,k=1$}\\
There is no solution for a closed model with zero lambda.
\item[V.] {\bf $\Lambda=0, k=0$}\\
The flat model with zero lambda presents the well-known Minkowski space.
\item[VI.] {\bf $\Lambda=0, k=-1$}\\
The open model with zero lambda presents Milne space.
\item[VII.] {\bf $\Lambda<0,k=+1$}\\
There is no solution for closed models with negative lambda.
\item[VIII.] {\bf $\Lambda<0,k=0$}\\
There is no solution for flat models with negative lambda.
\item[IX.] {\bf $\Lambda<0,k=-1$}\\
In this case eq.(\ref{d28}) has a negative lambda has a negative
discriminant which means there are three real roots.
Since this does not give a unique solution and the definition of function
is not fulfilled,
so should be rejected.
\end{itemize}

\section{Concluding Remarks}
\begin{itemize}
\item[1-] There exists no solution with oscillating nature.
\item[2-] Models $II$ and $III$ are singular and the nature of their singularities 
could not be explored by constructing a scalar of Riemann tensor.
It is illustrative to compute the density scalar $R$ by applying
eqs.(\ref{d25}) and (\ref{d26}),we get
\beq\label{f1}
R=-\frac 34\frac{r\sin^{\frac 12}\theta}{(1-kr^2)^{\frac 14}}\Lambda
\eeq
The action may be rewritten by inserting eq.(\ref{f1}) into eq.(\ref{a15})
\beq\label{f2}
I=\int{\kappa \frac{r^2\sin\theta}{\sqrt{1-kr^2}}(\frac 9{16}\Lambda^2)dt dr d\theta d\phi}
\eeq
This does reminisce a non-flat geometry with constant curvature.
So we may hope with some certain that singularity is non-intrinsic.
\item[3-] The Riemann tensor made of the metric (\ref{d2})
while satisfying (\ref{d27}) is consistent with eq.(\ref{c6}). So we may 
say with confidence that the obtained space-times are maximally symmetric.
\item[4-] Open models admit merely $\Lambda=0$.
\item[5-] The inflationary behavior of deSitter models are replaced by a 
power law character which reserve to be physically interpreted.
\item[6-] The event horizon $r_H$ given by eq.(\ref{e18})goes to infinity if
$t_0$ approaches to zero. We may hope while we have improved the
inflationary scenario also have resolved the existence of event horizon.   
\end{itemize}

\appendix
\section{1}
Non-zero components of the connection are:
\vspace{0.1em}
%\onecolumngrid
%\begin{widetext}
\beqa \label{12}
\Gamma^t_{tt}&=&\frac 14(\Gamma^\rho_{\rho t}-3\frac{\dot{a}}{a})\;\; ,\;\; \Gamma^t_{tr}=\frac 14(\Gamma^\rho_{\rho r}-\frac 2r-\frac {kr}{1-kr^2})\;\; , \;\;
\Gamma^t_{t\theta}=\frac 14(\Gamma^\rho_{\rho\theta}-\cot\theta)\;\;,\;\; \Gamma^t_{t\phi}=
\frac 14\Gamma^\rho_{\rho\phi} \nonumber\\
\Gamma^t_{rr}&=&\frac{a^2}{4(1-kr^2)}(\Gamma^\rho_{\rho t}
+\frac{\dot{a}}{a})\;\;,\;\;\Gamma^t_{\theta\theta}=\frac{a^2 r^2}{4}(\Gamma^\rho_{\rho t}+\frac{\dot{a}}{a})\;\;,\;\;\Gamma^t_{\phi\phi}=\frac {a^2r^2\sin^2\theta}{4}(\Gamma^\rho_{\rho t}+\frac{\dot{a}}{a}) \nonumber\\
\Gamma^r_{tt}&=&\frac{1-kr^2}{4a^2}(\Gamma^\rho_{\rho r}
-\frac 2r)-\frac{kr}{4a^2}\;\;,\;\;\Gamma^r_{rt}=\frac 14(\Gamma^\rho_{\rho t}+\frac{\dot a}{a})\;\;,\;\;\Gamma^r_{rr}=\frac 14(\Gamma^\rho_{\rho r}-\frac 2r+\frac{3kr}{1-kr^2})\nonumber\\
\Gamma^r_{r\theta}&=&\frac 14(\Gamma^\rho_{\rho r}-\cot\theta)\;\;,\;\;\Gamma^r_{r\phi}=\frac 14\Gamma^\rho_{\rho\phi}\;\;,\;\;\Gamma^r_{\theta\theta}=
-\frac 14 r^2(1-kr^2)(\Gamma^\rho_{\rho r}+\frac 2r)+\frac 14 kr^3\;\;,\;\;\Gamma^r_{\phi\phi}=\sin^2\theta\Gamma^r_{\theta\theta} \nonumber\\
\Gamma^\theta_{tt}&=&-\frac{1}{4a^2r^2}(\Gamma^\rho_{\rho\theta}-\cot\theta)\;\;,\;\;\Gamma^\theta_{\theta t}=\frac 14(\Gamma^\rho_{\rho t}+\frac{\dot a}{a})\;\;,\;\;\Gamma^\theta_{rr}
=-\frac 1{4r^2(1-kr^2)}(\Gamma^\rho_{\rho\theta}-\cot\theta)
\nonumber\\
\Gamma^\theta_{\theta r}&=&\frac 14(\Gamma^\rho_{\rho r}+\frac 2r-
\frac{kr}{1-kr^2})\;\;,\;\;\Gamma^\theta_{\theta\theta}=\frac 14
(\Gamma^\rho_{\rho\theta}-\cot\theta)\;,\;\Gamma^\theta_{\theta\phi}=\frac 14\Gamma^\rho_{\rho\phi}\;,\;\Gamma^\theta_{\phi\phi}
=-\frac{\sin^2\theta}{4}(\Gamma^\rho_{\rho\theta}+3\cot\theta)\nonumber\\
\Gamma^\phi_{tt}&=&\frac{\Gamma^\rho_{\rho\phi}}{4a^2r^2\sin^2\theta}
\;,\;\Gamma^\phi_{\phi t}=\frac 14(\Gamma^\rho_{\rho t}+\frac{\dot a}{a})\;,\;\Gamma^\phi_{rr}=-\frac{\Gamma^\rho_{\rho\phi}}{4r^2(1-kr^2)\sin^2\theta}\;\;,\;\Gamma^\phi_{\phi r}=\frac 14
(\Gamma^\rho_{\rho r}+\frac 2r-\frac{kr}{1-kr^2})\nonumber\\
\Gamma^\phi_{\theta\theta}&=&-\frac{\Gamma^\rho_{\rho\phi}}{4\sin^2\theta}\;,\;\Gamma^\phi_{\phi\theta}=\frac 14(\Gamma^\rho_{\rho\theta}+3\cot\theta)\;\;,\;\;\Gamma^\phi_{\phi\phi}
=\frac 14\Gamma^\rho_{\rho\phi} .\nonumber
\eeqa
%\end{widetext}
%\twocolumngrid
($\cdot$) represents differentiation with respect to time.

\appendix
\section{2}
The non-zero components of the connection after inserting eqs.(\ref{d6}),(\ref{d8})
and (\ref{d10}) in the given components of connection in A.1 are:
 \beqa 
 \Gamma^t_{tt}&=&\frac 14(\Gamma^\rho_{\rho t}-3\frac{\dot a}{a})\;\;,\;\;
 \Gamma^t_{rr}=\frac{a^2}{4(1-kr^2)}(\Gamma^\rho_{\rho t}+\frac{\dot a}{a})\;\;,\;\;\Gamma^t_{\theta\theta}=\frac 14 a^2r^2(\Gamma^\rho_{\rho t}+
 \frac{\dot a}{a})\;\;,\;\;\Gamma^t_{\phi\phi}=\Gamma^t_{\theta\theta}\sin^2\theta \n
 \Gamma^r_{rt}&=&\frac 14(\Gamma^\rho_{\rho t}+\frac{\dot a}{a})\;\;,\;\;
 \Gamma^r_{rr}=\frac{kr}{1-kr^2}\;\;,\;\;\Gamma^r_{\theta\theta}=-r(1-kr^2)
 \;\;,\;\;\Gamma^r_{\phi\phi}=\Gamma^r_{\theta\theta}\sin^2\theta \n
 \Gamma^\theta_{\theta t}&=&\frac 14(\Gamma^\rho_{\rho t}+\frac{\dot a}{a})\;\;,\;\;\Gamma^\theta_{\theta r}=\frac 1r\;\;,\;\;\Gamma^\theta_{\phi\phi}
 =-\sin\theta\cos\theta \n
 \Gamma^\phi_{\phi t}&=&\frac 14(\Gamma^\rho_{\rho t}+\frac{\dot a}{a})\;\;,\;\;
 \Gamma^\phi_{\phi r}=\frac 1r \;\;,\;\;\Gamma^\phi_{\phi\theta}=\cot\theta .\nonumber
 \eeqa 
%\acknowledgments
 
\end{document}